# Computational up-scaling of anisotropic swelling and mechanical behavior of hierarchical cellular materials

Ahmad Rafsanjani[1,2], Dominique Derome[1], Falk K. Wittel[3], Jan Carmeliet[1,4]

[1] *Laboratory for Building Science and Technology, Swiss Federal Laboratories for Materials Science and Technology, EMPA, Überlandstrasse 129, CH-8600 Dübendorf, Switzerland*

[2] *Department of Civil, Environmental and Geomatic Engineering, ETH Zurich, CH-8093 Zurich, Switzerland*

[3] *Computational Physics for Engineering Materials, IfB, HIF E12, ETH Zurich, Schafmattstr. 6, CH-8093 Zurich, Switzerland*

[4] *Chair of Building Physics, ETH Zurich, HIL E46.3, Wolfgang-Pauli-strasse 15, CH-8093 Zurich, Switzerland*

**Abstract**

The hygro-mechanical behavior of a hierarchical cellular material, i.e. growth rings of softwood is investigated using a two-scale micro-mechanics model based on a computational homogenization technique. The lower scale considers the individual wood cells of varying geometry and dimensions. Honeycomb unit cells with periodic boundary conditions are utilized to calculate the mechanical properties and swelling coefficients of wood cells. Using the cellular scale results, the anisotropy in mechanical and swelling behavior of a growth ring in transverse directions is investigated. Predicted results are found to be comparable to experimental data. It is found that the orthotropic swelling properties of the cell wall in thin-walled earlywood cells produce anisotropic swelling behavior while, in thick latewood cells, this anisotropy vanishes. The proposed approach provides the ability to consider the complex microstructure when predicting the effective mechanical and swelling properties of softwood.

**Keywords**: A. Wood, B. Mechanical properties, C. Multiscale modeling, C. Anisotropy, Homogenization

## 1. Introduction

Many natural biological materials show a hierarchical structure which determines their mechanical behavior [1]. The swelling and shrinkage due to wetting or drying are not fully reversible and also highly determined by the hierarchical structure of the material. In this paper, we study wood, a natural composite material of multiscale configuration. In the case of temperate climate species like Norway spruce, the hierarchy in cellular structure comes from the seasonal growth. The complicated microstructural architecture of wood, as shown in Figure 1, introduces a strong geometric anisotropy which is reflected in the anisotropy of its mechanical and swelling behavior. A better understanding and predictability of the interactions between the mechanical and moisture behavior of wood and its dependency on the cellular architecture is needed, for example, to assess the durability of wood elements exposed to varying mechanical and environmental loading.

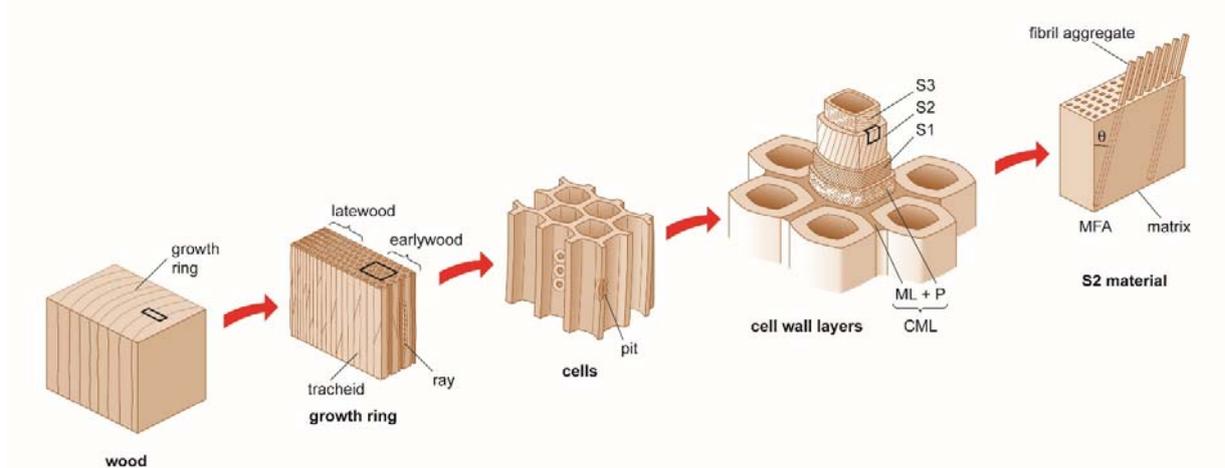

**Figure 1**. Multiscale structure of wood

The interaction of the moisture and mechanical behavior of wood is best observed in swelling. Absorption of moisture in wood, in the hygroscopic range i.e. until around 30% moisture content mass per mass, results in swelling and reduced stiffness. The microscopic origin of this behavior can be found on the level of the cell wall. The cell wall material is composed in almost equal quantity of stiff cellulose microfibrils and a soft polymeric matrix (see Figure 1). The hydrophobic crystalline cellulose is surrounded by hydrophilic amorphous cellulose, immersed in a hydrophilic amorphous matrix of hemicelluloses bound by lignin. Sorption of water molecules in between hydrophilic molecules pushes the constituents apart, resulting in swelling and a reduction of stiffness of the matrix. As represented in Figure 1, the thin internal and external cell wall layers (namely S3 and S1) act as corsets due to the winding of the cellulose fibrils around the cell. In the central and, by far, thickest cell wall layer, namely S2, the cellulose microfibrils are almost parallel to the longitudinal axis of the cells, although the presence of an angle (called microfibril angle, MFA) results in a helicoidal organization of the fibrils. Although the effect of varying the MFA from 0° to 30° on the transverse stiffness and swelling properties of softwoods is small during the sorption of moisture [2], the general orientation of the microfibrils in the S2 layer results in notable swelling in the transverse directions of the cell and almost none along the longitudinal (L) direction [3]. Furthermore, the helicoidal organization of the cellulose fibrils in the cell wall causes swelling to be more important normal to the cell wall than along the cell wall direction [4]. Although swelling originates from the cell wall level, its anisotropic nature finds its origin mainly at the cellular architecture level. On a mesoscopic scale, wood consists mainly of longitudinal tracheid cells and radially oriented ray cells, the latter representing 5% of the wood volume. Across the growth ring, the thin-walled earlywood cells with large internal cavities, called lumens, gradually change to thick-walled latewood cells with small-sized lumens. The variation of cellular structure across the growth ring is found responsible for the anisotropy of swelling and stiffness [5, 6]. Furthermore, findings on the isotropic swelling of isolated latewood and anisotropic swelling of isolated earlywood in the transverse plane indicate the possibility of latewood/earlywood interaction [7].

To study wood, different cellular models have been developed using regular/irregular hexagonal honeycombs with isotropic cell wall material properties [6, 8]. These models can predict the elastic behavior of wood, however they are unable to capture differential transverse swelling [9]. In multiscale approaches, the homogenization theory has been commonly used to span over several scales, from the cell wall to the timber scale, with intermediary levels like the cell and the growth ring [10-12]. Swelling anisotropy, in particular, is rarely considered in computational models. The primary purpose of this work is to study the anisotropic hygro-mechanical properties of hierarchical cellular materials using a multiscale framework. The growth ring behavior is easily accessible and is considered here as the macro-scale for wood. We investigate the distribution of transverse anisotropy in swelling and mechanical behavior of wood from the cellular scale to the growth ring level by means of a finite element-based multi-scale approach and comparison of the numerical results with experiments. The mechanical fields at the macroscopic level (growth ring) is resolved through the incorporation of the micro-structural (cellular structure) response by the computational homogenization of different Representative Volume Element (RVE) of wood cells selected from a morphological analysis of wood at the cellular scale. The proposed multiscale framework aims at achieving a more realistic characterization of the anisotropic swelling and mechanical behavior of wood and, in particular, at describing the latewood and earlywood interaction caused by the material heterogeneity on the cellular scale.

## 2. Experimental work

### 2.1 Swelling of earlywood and latewood specimens

High-resolution phase contrast synchrotron X-ray tomography measurements of the free swelling of isolated earlywood and latewood specimens exposed to varying relative humidity conditions were conducted [7]. In a dynamic vapor sorption (DVS) apparatus, the samples were subjected to the same relative humidity protocol to obtain the moisture content. Figure 2a shows the sorption isotherm for earlywood and latewood during the adsorption from 25% to 85% relative humidity (RH). The measurements used to determine swelling coefficients for early- and latewood are illustrated in Figure 2b and 2c respectively. Latewood shows higher swelling strains than earlywood and rather isotropic cellular deformations, which means that the cells keep their initial shape. For earlywood, the free swelling strain in the radial direction is more than 3 times less than the one in the tangential direction while for the latewood this ratio is close to 1. The linear swelling coefficients (%strain/%moisture content) were measured in radial and tangential directions to be $\beta_R = 0.065$ and $\beta_T = 0.207$ for earlywood and $\beta_R = 0.300$ and $\beta_T = 0.347$ for latewood samples.

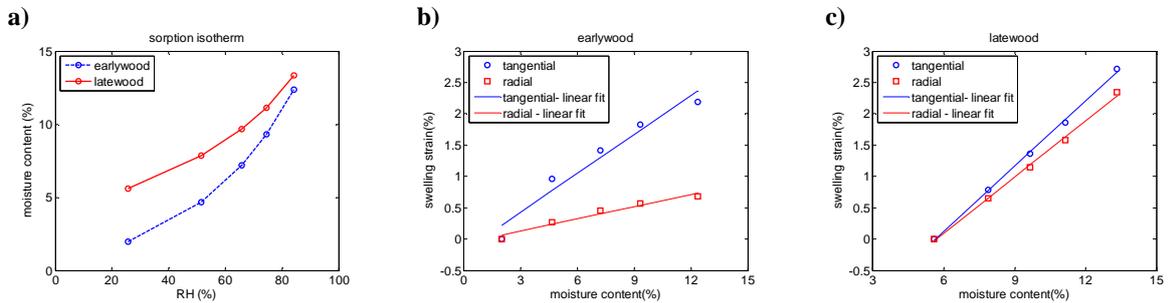

**Figure 2.** (a) Sorption isotherm for earlywood and latewood and (b) swelling strain of earlywood and (c) latewood in radial and tangential directions against moisture content.

*2.2 Swelling of growth rings*

The swelling of a growth ring was analyzed experimentally using environmental scanning electron microscopy (ESEM), on images of a growth ring taken from a larger piece of wood, at two different relative humidities, i.e. 32% and 64%RH. Figure 3a shows the reference SEM image at 32%RH. The measured displacement are analyzed based on the affine transformation, $u_{aff} = x' - x = (F-1)x + c,$ where $F$ is the affine transformation matrix composed of $F = S \cdot L \cdot R$, with $S$ the shear matrix, $L$ the scaling matrix and R the rotation matrix. This matrix is used to register the image and to determine the remaining displacements, i.e. the non-affine displacements. Assuming further that the non-affine displacements scale linearly with the moisture content variation, as is assumed for the affine displacements, we obtain the displacement normalized to the moisture content. Since it was not possible to determine the exact moisture content of the sample in the ESEM chamber, the displacements were scaled to the maximum (negative) displacement and are presented in Figure 3c along the relative ring position (0 marks the start of the growth ring and 1 the end). According to this registration procedure, a non-zero non-affine displacement indicates non-homogenous cellular deformation (see Figure 3b). We observe that the highest displacements occur in radial direction in the middle of the growth ring, i.e. in a transition zone between earlywood and latewood, where the more important swelling of the stiffer latewood pushes radially the softer earlywood. The cell swelling behavior in radial direction along the growth ring is thus significantly different between earlywood and latewood while in tangential direction is almost homogeneous for earlywood and latewood.

a)

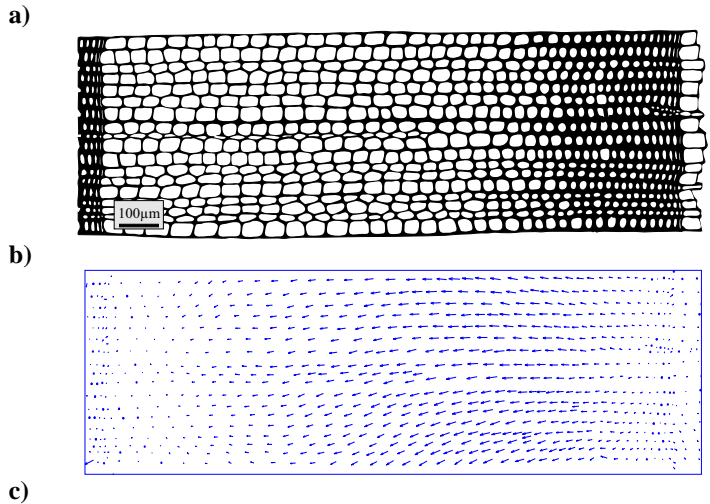

b)

c)

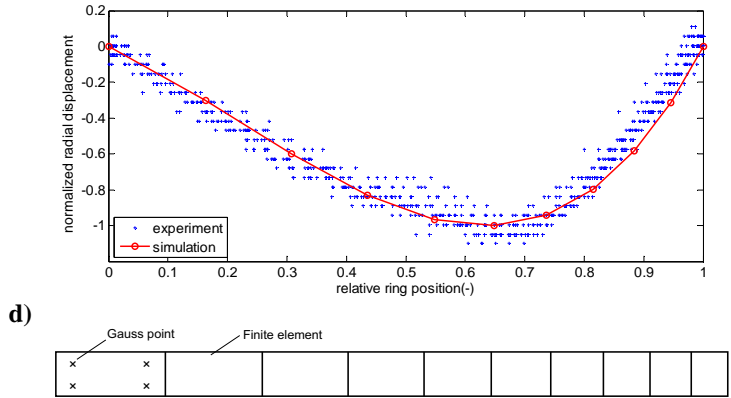

**Figure 3.** Growth ring of Norway spruce wood. (a) Binary image of ESEM micrograph of radial-tangential cross section and (b) deformation field between 32% and 64% relative humidity. (c) Experimental and simulation of normalized radial displacement during swelling. (d) Macroscopic FEM mesh.

## 3. Modeling procedure

We consider a problem of infinitesimal small deformations with moisture induced swelling for growth rings in softwood as a cellular porous solid within the framework of the computational homogenization. Since the longitudinal dimensions of the wood cells are very long in comparison to the dimensions in radial and tangential directions, the problem can be reduced to the analysis of the cross-section of the material, which is in a state of generalized plane strain. Verifying for separation of scales, the characteristic size of wood cells (micro-scale) is much smaller (~40 $\mu m$) than the growth ring (macro-scale) length scale (~3 $mm$). In the regions far away from the center of the stem of the tree, growth rings are periodically arranged in radial direction which justifies an assumption on global periodicity of the timber. At the micro-scale, we assume local periodicity, where, although wood cells have different morphologies corresponding to different relative ring positions, each cell repeats itself in its vicinity.

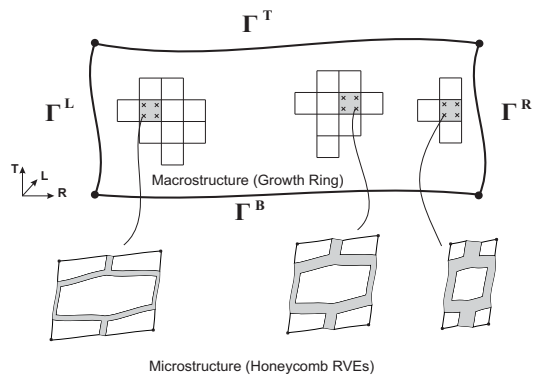

**Figure 4.** Two-scale upscaling scheme and underlying RVEs corresponding to different elements

As a most probable assumption, honeycomb RVEs are selected for this current analysis. The concept of local and global periodicity in softwood and the proposed two-scale upscaling scheme is illustrated in Figure 4. It should be also mentioned that although ray cells may have a stiffening effect [13] and a limiting effect on swelling [7] in radial direction, they are excluded for now in this first two-dimensional application of the homogenization approach proposed in the paper. Material properties of the cell wall are taken from literature [4, 12] and considered for the moment to be independent of moisture content.

*3.1 Governing equations*

At the micro-scale (subscript m), the equilibrium equation for stress $\sigma_m$ and the constitutive equation of a linearly elastic solid (cell wall) with moisture induced swelling are respectively given as follows:

$$\nabla_m . \sigma_m = \vec{0}, \quad in\ \Omega_m \tag{1}$$

$$\sigma_{ij}^m = C_{ijkl}^m (\varepsilon_{kl}^m - \beta_{kl}^m \Delta m) \tag{2}$$

where $\Omega_m$ is the volume, $C_{ijkl}^m$ is the microscopic elasticity tensor, and $\nabla$ the symmetric gradient operator. Here, $\Delta m$ is the change in moisture content by mass from the initial state and $\beta_{kl}^m$ is the microscopic second order tensor of swelling coefficients. At the macro level (subscript M), the mechanical equilibrium and the constitutive equation have structures identical to the micro level,

$$\nabla_M . \sigma_M = \vec{0}, \tag{3}$$

$$\sigma_{ij}^M = C_{ijkl}^M (\varepsilon_{kl}^M - \beta_{kl}^M \Delta m) \tag{4}$$

which are complemented by macroscopic mechanical boundary conditions. The macroscopic constitutive behavior is obtained from the solution of the micro-scale problem defined on the underlying RVE as elaborated in the following.

*3.2 Upscaling approach*

The macroscopic strain tensor at any arbitrary point of the macroscopic continuum is achieved by volume averaging of the microscopic strain tensor field over the domain $\Omega_m$:

$$\varepsilon_M = \frac{1}{\Omega_m} \int_{\Omega_m} \varepsilon_m \, d\Omega. \tag{5}$$

Furthermore, it is possible to decompose the displacement field at a location $\vec{x}$ within RVE into a macroscopic linear displacement contribution and a displacement fluctuation:

$$\vec{u}_m(\vec{x}) = \varepsilon_M . \vec{x} + \vec{u}_f(\vec{x}). \tag{6}$$

The displacement fluctuations field ($\vec{u}_f$) represents the fine scale deviations with respect to the average fields as a result of the heterogeneities within the RVE. The micro-to-macro transition is essentially based on the Hill condition [14] which guarantees the equivalence between the energetically and mechanically defined effective elastic properties:

$$\sigma_M : \varepsilon_M = \frac{1}{\Omega_m} \int_{\Omega_m} \sigma_m : \varepsilon_m \, d\Omega, \qquad (7)$$

and is the basis for the different types of boundary conditions that can be imposed at the micro level. In the case of periodic microstructures, the periodic boundary conditions can be utilized and are known to yield more accurate estimation of apparent macroscopic properties [15]. For periodic boundary conditions, the boundary must appear in parallel pairs of corresponding surfaces, i.e. $(\Gamma^r, \Gamma^l)$ and $(\Gamma^t, \Gamma^b)$ as depicted in Figure 5a, where subscripts $l$, $r$, $b$ and $t$ refer to surfaces on the left, right, bottom and top boundary of the RVE, respectively.

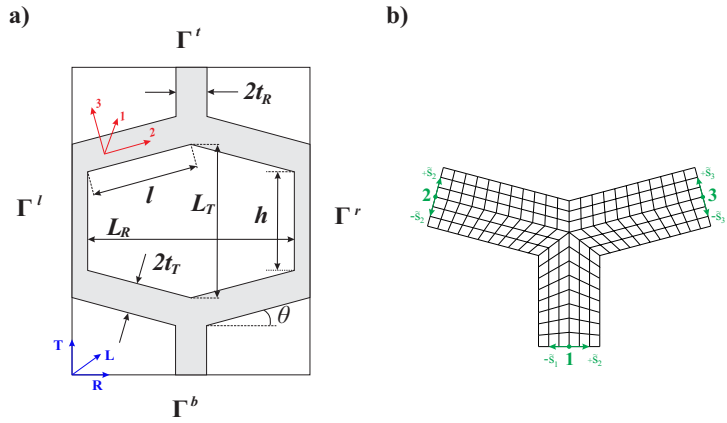

**Figure 5.** Honeycomb RVE (a) geometrical parameters (b) discretized quarter of a RVE.

The implementation of periodic boundary conditions leads to the following relations:

$$\vec{u}_r - \vec{u}_l = \varepsilon_M \cdot \Delta x_{lr}, \quad \vec{u}_t - \vec{u}_b = \varepsilon_M \cdot \Delta x_{bt}, \qquad (8a,b)$$

where $\Delta x_{lr}$ and $\Delta x_{bt}$ are constant distance vectors between corresponding surfaces. The macroscopic stresses can be computed numerically in a classical manner:

$$\sigma_M = \frac{1}{\Omega_m} \int_{\Omega_m} \sigma_m \, d\Omega = \frac{1}{\Omega_m} \sum_{n_{IP}} \sigma_m \, \Omega_{IP}, \qquad (9)$$

where $n_{IP}$ is the number of integration points (Gauss points) and $\Omega_{IP}$ is the integration point volume. The stress averaging procedure is computationally expensive. To compute the effective elastic stiffness tensor, an alternative approach is the master node technique [16]. In this method, the consistent macroscopic stiffness matrix at a macroscopic integration point is derived from the total RVE stiffness matrix by reducing the latter to the relation between forces acting on the retained vertices of the RVE and the displacements of these vertices. Thus, only a few resultant forces and displacements have to be read and post-processed from the finite element database.

Using this approach, the extraction of effective macroscopic stiffness is reduced to one analysis instead of four successive load cases ($\varepsilon_{11}, \varepsilon_{22}, \varepsilon_{12}$ and $\varepsilon_{33}$). Furthermore, we note that the honeycomb-RVE under periodic boundary conditions satisfies both translational and point symmetry. Thus, in our case, the analysis can be reduced to a symmetric quarter of the original RVE as depicted in Figure 5b. It is shown in [17] that substantial savings in computing times are achieved by taking advantage of such symmetries. These symmetry conditions imply that the displacement fluctuations of the pivot points 1, 2 and 3 (see Figure 5b) vanish as expressed in:

$$\vec{u}_f(\vec{x}_i) = \vec{0}, \; i = 1, \ldots, 3. \tag{10}$$

These pivot points, plus a fictitious node (node 0) corresponding to the contribution of the uniform out-of-plane deformation shared by all the elements forming the mesh, are selected as master nodes in this analysis. As a consequence of zero displacement fluctuation (10), the displacement vectors of master nodes using relation (6) can be written as follows:

$$\vec{u}_p = \boldsymbol{\varepsilon}_M \cdot \vec{x}_p, \quad p = 1, 2, 3, 0. \tag{11}$$

The symmetric quarter of the RVE is discretized in such a way that the nodes on antisymmetric pairs match geometrically. This eases the implementation of the periodicity condition through the following constrained relations

$$\vec{x}(-\tilde{s}_p) + \vec{x}(+\tilde{s}_p) = 2\vec{x}_p, \quad p = 1, 2, 3, \tag{12}$$

where $\tilde{s}_p$ denotes a local coordinate system centered on pivot points $p = 1, 2, 3$ [18]. The application of the averaging theorem on $\boldsymbol{\sigma}_m$ over $\Omega_m$, and the transformation from volume integral (9) to surface integral can be elaborated as follows

$$\boldsymbol{\sigma}_M = \frac{1}{\Omega_m} \int_\Gamma \vec{t} \otimes \vec{x} \, d\Gamma, \tag{13}$$

where $\Gamma$ is the boundary of RVE and $\vec{t} = \vec{n} \cdot \boldsymbol{\sigma}_m$ is the Cauchy stress vector with the normal vector $\vec{n}$. The symbol $\otimes$ denotes the dyadic product. Therefore, it is computationally more efficient to compute $\boldsymbol{\sigma}_M$ through the integral (13). In the case of periodic boundary conditions, $\boldsymbol{\sigma}_M$ can be further simplified as

$$\boldsymbol{\sigma}_M = \frac{1}{\Omega_m} \sum_{p=0,1,2,3} \vec{f}_p^e \otimes \vec{x}_p, \tag{14}$$

where $\vec{f}_p^e$ is the reaction external force acting on the master nodes. The macroscopic tangent stiffness is obtained numerically from the relation between the variation of the macroscopic stress and the variation of the macroscopic deformation at every macroscopic integration point. For extraction of macroscopic stiffness, the total system of equations for the RVE is rearranged as

$$\begin{bmatrix} K_{dd} & K_{dp} \\ K_{pd} & K_{pp} \end{bmatrix} \begin{bmatrix} \delta u_d \\ \delta u_p \end{bmatrix} = \begin{bmatrix} 0 \\ \delta f_p \end{bmatrix}, \qquad (15)$$

with $K_{dd}, K_{dp}, K_{pd}$ and $K_{pp}$ partitions of the stiffness matrix while $\delta u_p$ and $\delta f_p$ refer respectively to the incremental displacements and external forces of the prescribed retained vertices and $\delta u_d$ to the incremental displacements of the dependent nodes. By condensing out the dependent degrees of freedom from the system, the reduced stiffness matrix $\boldsymbol{K}_M$ is obtained as

$$K_M \delta u_p = \delta f_p \text{ with } K_M = K_{pp} - K_{pd} K_{dd}^{-1} K_{dp}. \qquad (16)$$

For the generalized plane strain assumption, $\boldsymbol{K}_M$ is a $[7 \times 7]$ matrix where the last row and column correspond to the contribution of the uniform out-of-plane deformation:

$$\boldsymbol{K}_M = \begin{bmatrix} \begin{bmatrix} K_{11}^{11} & K_{12}^{11} \\ K_{21}^{11} & K_{22}^{11} \end{bmatrix} & \begin{bmatrix} K_{11}^{12} & K_{12}^{12} \\ K_{21}^{12} & K_{22}^{12} \end{bmatrix} & \begin{bmatrix} K_{11}^{13} & K_{12}^{13} \\ K_{21}^{13} & K_{22}^{13} \end{bmatrix} & \begin{bmatrix} K_{13}^{10} \\ K_{23}^{10} \end{bmatrix} \\ \begin{bmatrix} K_{11}^{21} & K_{12}^{21} \\ K_{21}^{21} & K_{22}^{21} \end{bmatrix} & \begin{bmatrix} K_{11}^{22} & K_{12}^{22} \\ K_{21}^{22} & K_{22}^{22} \end{bmatrix} & \begin{bmatrix} K_{11}^{23} & K_{12}^{23} \\ K_{21}^{24} & K_{22}^{24} \end{bmatrix} & \begin{bmatrix} K_{13}^{20} \\ K_{23}^{20} \end{bmatrix} \\ \begin{bmatrix} K_{11}^{31} & K_{12}^{31} \\ K_{21}^{31} & K_{22}^{31} \end{bmatrix} & \begin{bmatrix} K_{11}^{32} & K_{12}^{32} \\ K_{21}^{32} & K_{22}^{32} \end{bmatrix} & \begin{bmatrix} K_{11}^{33} & K_{12}^{33} \\ K_{21}^{33} & K_{22}^{33} \end{bmatrix} & \begin{bmatrix} K_{13}^{30} \\ K_{23}^{30} \end{bmatrix} \\ \begin{bmatrix} K_{31}^{01} & K_{32}^{01} \end{bmatrix} & \begin{bmatrix} K_{31}^{02} & K_{32}^{02} \end{bmatrix} & \begin{bmatrix} K_{31}^{03} & K_{32}^{03} \end{bmatrix} & \begin{bmatrix} K_{33}^{00} \end{bmatrix} \end{bmatrix}. \qquad (17)$$

Once the reduced stiffness matrix is obtained, the consistent tangent stiffness can be derived as (for the derivation, see [15]):

$$\boldsymbol{C}_M = \frac{1}{\Omega_m} \sum_i \sum_j \left( \vec{x}_i K_M^{ij} \vec{x}_j \right)^{LC}, \; i,j = 1,2,3,0, \qquad (18)$$

where $K_M^{ij}$ is a sub-matrix of the matrix $\boldsymbol{K}_M$, at the rows and columns of the degree of freedom in the nodes $i$ and $j$. $\vec{x}_i$ and $\vec{x}_j$ are respectively the position vectors of these nodes. The left conjugation of a fourth order tensor is denoted by $(.)^{LC}$ and defined as $T_{ijkl}^{LC} = T_{jikl}$ in index notation. The macroscopic swelling coefficients, $\beta_{ij}^M$, can be calculated by adding an auxiliary load case that constrains all displacements normal to the RVE boundaries (i.e., $\boldsymbol{\varepsilon}_M = \boldsymbol{0}$), and applies a unit moisture content increment ($\Delta m = 1$). This allows evaluating the macroscopic swelling stress tensor $\sigma_{ij}^M$, using (14). Inserting $\sigma_{ij}^M$ in the macroscopic constitutive equation (4), the macroscopic swelling coefficients can be readily obtained:

$$\beta_{ij}^M = -C_{ijkl}^{M\,-1} \sigma_{kl}^M. \qquad (19)$$

*3.3 Cell geometry and cell wall properties*

The cell wall properties are assumed to be orthotropic and the three orthogonal material directions (see Figure 5a) are selected to be oriented normal to the plane (1-axis), along the cell wall (2-axis) and normal to the cell wall thickness (3-axis). The material properties of cell wall for earlywood and latewood are provided in Table 1.

**Table 1:** Elastic properties of the cell wall (in GPa) from micromechanical model [12] and cell wall swelling coefficients (% strain /% moisture content) from experiment [4]

| sample | $E_{11}$ | $E_{22}$ | $E_{33}$ | $G_{12}$ | $G_{13}$ | $G_{23}$ | $\nu_{21}$ | $\nu_{31}$ | $\nu_{32}$ |
|---|---|---|---|---|---|---|---|---|---|
| earlywood | 33.20 | 7.02 | 4.36 | 4.38 | 1.65 | 1.18 | 0.112 | 0.023 | 0.403 |
| latewood | 43.00 | 6.43 | 4.77 | 3.50 | 2.12 | 1.22 | 0.067 | 0.022 | 0.416 |

| | $\beta_{22}$ | | $\beta_{33}$ | |
|---|---|---|---|---|
| | Radial wall | Tangential wall | Radial wall | Tangential wall |
| earlywood | 0.10 | 0.15 | 0.45 | 0.45 |
| latewood | 0.40 | 0.35 | 0.50 | 0.60 |

The elastic properties are chosen from [12] and the swelling coefficients are selected from [4]. These values should be considered as the effective material properties of the whole cell wall. The dimensions of the cellular structure were characterized using an SEM image of a single growth ring of Norway spruce (Picea abies) on 10 continuous cell rows in radial direction [19]. The lumen diameters and the cell wall thickness along the relative ring position and the corresponding fitted functions are presented in Figures 6a and b respectively. These values in combination with a shape angle $\theta$ (see Figure 5a) are used for indirect characterization of the remaining parameters (i. e. $h$ and $l$) of the honeycomb RVEs. In this analysis, we quantified the shape angle parameter to range linearly from $\theta \approx 5°$ in earlywood to $\theta \approx 10°$ in transition zone and then increased it linearly up to $\theta \approx 20°$ in latewood. The density of each cell can be calculated from the volume fraction of the cell wall in an RVE multiplied by the cell wall density $1466\,kg/m^3$ [19].

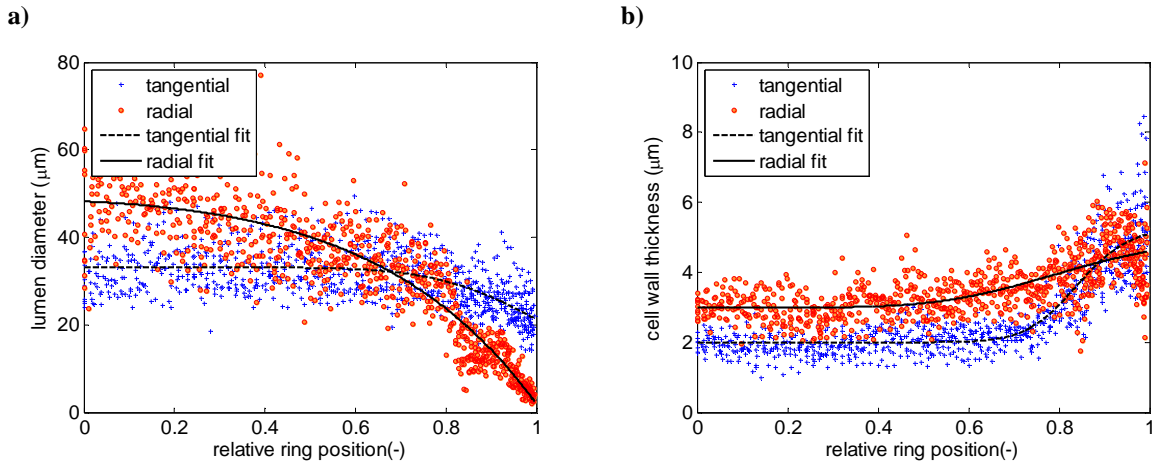

**Figure 6.** (a) Lumen diameter and (b) cell wall thickness in radial and tangential directions as a function of the relative ring position.

*3.4 Numerical implementation*

The two-scale hygro-mechanical framework has been implemented in the finite element package ABAQUS using user-defined material routines (ABAQUS Inc., Providence, RI). According to the distribution of earlywood and latewood in a specific growth ring, the macro domain is discretized. Since the variation in geometrical parameters in tangential direction is negligible, the macro domain is meshed only in radial direction using standard four-node elements with finer resolution for latewood. The macroscopic mesh is shown in Figure 3d. In this study, material nonlinearities are not considered which makes the implementation of two-scale analysis simple. As the geometrical parameters of wood cells only depends on relative ring position, all macroscopic stiffness and swelling tensors can be determined prior to the macroscopic analysis. The homogenized material properties can be evaluated and stored for each integration point in the macro domain or by interpolated functions obtained from the analysis of micro-scale problems. Here, the upscaling procedure is carried out in two steps. First, at each integration point, based on the micro-structural geometrical distribution of wood cells along the relative growth ring position, a honeycomb unit cell is generated. The reduced stiffness matrix of the RVEs at the retained master nodes is extracted and used in Eq. 18 to calculate the homogenized stiffness tensor. In the second step, the micro-scale analysis is followed by a mechanical equilibrium in which a uniform moisture increment is applied to a constrained unit cell. The average resultant stresses in the RVE are calculated using Eq. 14 and, together with the computed effective stiffness tensor, are inserted into Eq. 19 for calculating the effective swelling coefficients. Finally, the stored effective properties are transferred to the macro-problem.

## 4. Results and discussion

To better understand the swelling and mechanical behavior of softwoods, different parameter studies at cellular scale are performed. Then, the two-scale model is utilized to study the behavior of a single growth ring. The simulated results are compared to experiments at both scales.

*5.1 Unit cell*

We investigate the effects of the unit cell geometry and of the cell wall material anisotropy on the elastic and swelling properties of individual wood cells. We assigned earlywood and latewood material properties to the start and end of the relative ring position respectively. Within the growth ring the material properties are assumed to be linearly dependent on the cell wall thickness, due to the increasing thickness of the S2 layer. The effective swelling coefficients in tangential (T) and radial (R) directions and the respective T/R anisotropy ratios are shown in Figure 7 and compared with microscopic observation of transverse swelling of isolated earlywood and latewood samples as explained in Section 2.1. The tangential swelling coefficients are generally greater than the radial ones, except in the region close to the end of the growth ring as presented in Figure 7a.

a)                                        b)

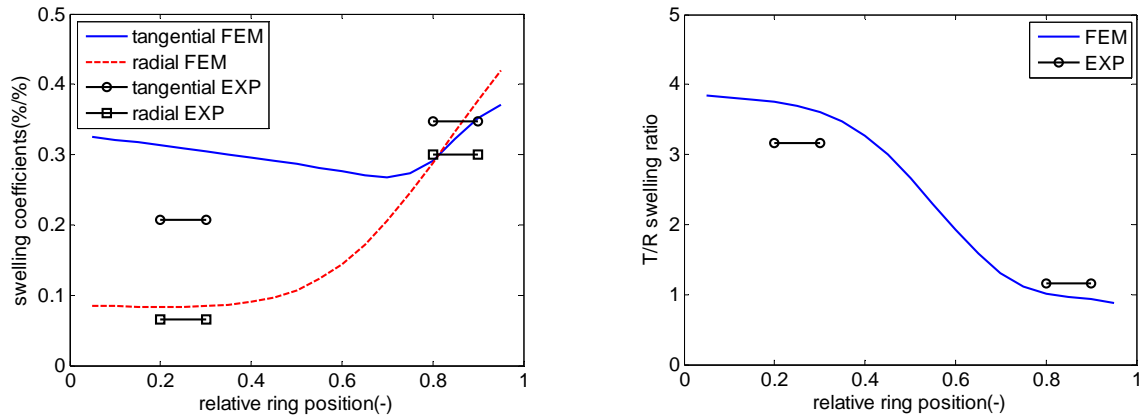

**Figure 7.** (a) The homogenized tangential and radial swelling coefficients for honeycomb unit cells and (b) the respective T/R swelling ratios and experimental results for Norway spruce [7].

The swelling coefficients of latewood cells are greater than of earlywood ones. The increase of swelling coefficients in radial direction is more significant than in tangential direction. The T/R swelling ratio in latewood is smaller than in earlywood and swelling in latewood is almost isotropic (value equal to 1). The swelling of earlywood cells is on the contrary clearly anisotropic, with swelling ratios up to 4. This observation is remarkable since the cell wall elastic properties of earlywood and latewood cells do not differ significantly which means that the main influence comes from the geometrical parameters and swelling coefficients of the cell wall. We observe good agreement between simulations and experiments for both earlywood and latewood tissues. Figure 8a gives the elastic modulus in tangential and radial directions along the growth ring and experiments of micro-tensile tests in the transverse plane of earlywood and latewood parts of spruce [20]. The elastic modulus in earlywood is very low and increases strongly in the latewood region. T/R elastic anisotropy ratios of honeycomb unit cells are depicted in Figure 8b. In earlywood region, the modulus of elasticity in radial direction is greater than in the tangential one, while the opposite is obtained in the latewood region. This observation can be explained considering the change of the geometry of earlywood and latewood cells over the growth ring. As it can be seen in Figure 6a, the radial lumen dimension $L_R$ gradually decreases over the earlywood layer, followed by an even more pronounced decrease in the latewood layer which provides considerably stiffer material properties in tangential direction. The lumen dimension in the tangential direction $L_T$ is almost constant throughout the year, with a slight decrease in the latewood layer. Therefore, less increase in radial elastic modulus is observed. In conclusion, our results show that the cell geometry and the orthotropic material properties of the cell wall are responsible for the anisotropy in stiffness and swelling properties of softwood cells. In particular, the different swelling coefficients within the cell wall in transverse plane play an important role in the anisotropic swelling of earlywood cells. These findings are in line with the rare experimental data available at the cell wall level [4] and lead us in distinguishing different swelling mechanism in earlywood and latewood.

a)          b)

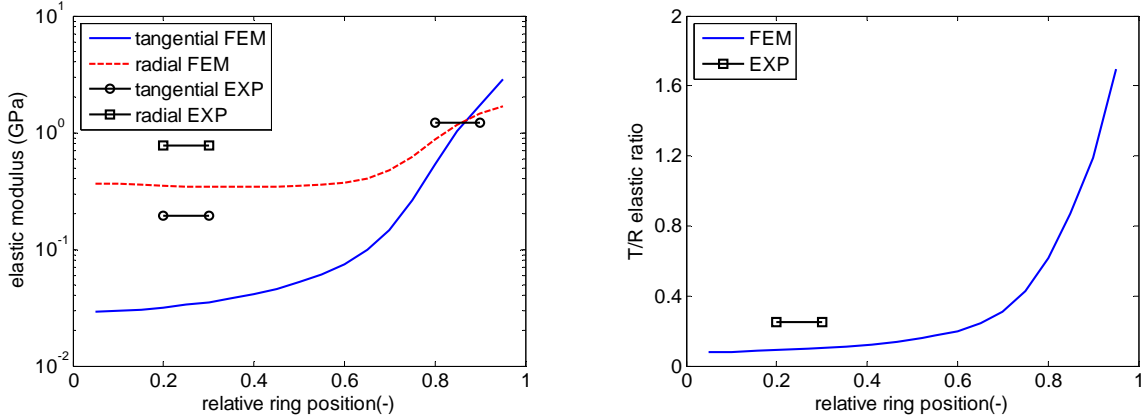

**Figure 8.** (a) The homogenized tangential and radial elastic moduli for honeycomb unit cells and (b) the respective T/R elastic ratios and experimental results for spruce [20].

*5.2 Growth ring*

The two-scale model is now used to calculate the effective properties of a specific growth ring. The geometrical parameters of the underlying RVEs are related to the relative ring position with the aid of the functions which are fitted to the measurements of the wood cell dimensions as depicted in Figure 6. The scatter of the geometrical parameters in tangential direction is ignored. This simplification is reasonable since the deformation pattern in tangential direction is almost uniform (see Figure 3b). The resulting effective elastic properties and swelling coefficients are presented and confronted with literature data in Table 2.

**Table 2:** Effective mechanical and swelling properties of growth rings in Spruce (density in $Kg/m^3$, elastic properties in GPa and swelling coefficients in %strain/% moisture content)

| sample | $\rho$ | $E_R$ | $E_T$ | $E_L$ | $G_{RT}$ | $\nu_{TR}$ | $\nu_{RL}$ | $\nu_{TL}$ | $\beta_R$ | $\beta_T$ | $\beta_L$ |
|---|---|---|---|---|---|---|---|---|---|---|---|
| present work | 443 | 0.757 | 0.460 | 11.13 | 0.030 | 0.34 | 0.035 | 0.017 | 0.17 | 0.31 | 0.009 |
| [21] | 390 | 0.680 | 0.430 | 10.80 | 0.030 | 0.31 | 0.022 | 0.019 | - | - | - |
| [22] | 460 | 0.625 | 0.397 | 12.80 | 0.053 | 0.21 | 0.018 | 0.014 | - | - | - |
| [23] | 375 | | | | | | | | 0.19 | 0.37 | - |
| [24] | 480 | | | | | | | | 0.15 | 0.32 | - |

The result of the growth ring simulation indicates that the tangential swelling is greater than the radial one. The periodic boundary condition which is imposed on the growth ring level includes the effect of the earlywood and latewood alternation for softwoods grown in the temperate zone. As a result, the strong bands of latewood in tangential direction force the weak bands of earlywood to swell tangentially to about the same extent as the latewood. Finally, we analyze whether the present multiscale model can predict the experimental trend at the growth ring level as explained in Section 2.2. The resulting experimental deformation field is compared to simulation results of a two-scale growth ring model and is presented in Figure 3c. The free swelling simulation is followed by the registration procedure as done in the processing of the SEM images. The comparison gives us an insight about the interaction of the earlywood and latewood tissues within the growth ring. A good agreement between experiment and simulation is obtained. As it can be seen, the maximum deformation in the simulation occurs also at the earlywood-latewood transition front towards the earlywood.

## 6. Conclusion

In this work, the hygro-mechanical behavior of growth rings is investigated using a two-scale model based on a computational homogenization technique taking both material anisotropy of the cell walls and the geometry of the cellular structure into account. This model is used to predict transverse anisotropy in mechanical and swelling properties of softwoods. Unit cell and two-scale finite element simulations for the growth ring are conducted to investigate the hygro-mechanical behavior of wood at respective hierarchical levels. Unit cell simulations show high tangential/radial (T/R) swelling ratio at the start of the growth ring in earlywood which decreases along the relative ring position and become close to 1.0 in latewood. The T/R elastic ratio shows an opposite trend which is almost constant in earlywood and increase significantly in latewood region. The orthotropic swelling properties of the cell wall material induce anisotropic dimensional change in thin-walled earlywood cells while, in thick latewood cells, this anisotropy vanishes. The predicted swelling coefficients are compared to experimental data of earlywood and latewood samples. The agreement is good, predicting general trends accurately and giving a physical understanding of the observed phenomena. In addition, the complex structure of a natural material, such as wood, introduces a superposition of different influences which makes it difficult to evaluate. Finally, one growth ring is modeled using the two-scale model. The results are in good agreement with existing values from literature. The framework has the flexibility to include different non-linear and moisture-dependent constitutive material behavior at the micro level. However, for the analysis of nonlinear moisture effects, the macroscopic equilibrium problems would have to be solved simultaneously with the microscopic equilibrium at Gauss-point level for all loading steps. The proposed model can be used for studying the influence of morphological, mechanical and moisture effects on effective hygro-mechanical properties, here, in softwoods and eventually in other wood-based composites.

**Acknowledgment**

The authors are grateful for the financial support of the Swiss National Science Foundation (SNF) under Grant No. 125184.